\newcommand{\PSD}{\hat{\psi}^\dagger}
\newcommand{\PSI}{\hat{\psi}}
\newcommand{\rv}{({\bf r})}
\newcommand{\D}{{\rm d}}

\documentstyle[twocolumn,epsf]{jpsj}

\title {
  Vortex Stabilization in Dilute Bose-Einstein Condensate Under 
Rotation
}

\author {
  Tomoya {\sc Isoshima}\footnote{E-mail: tomoya@mp.okayama-u.ac.jp}
   and Kazushige {\sc Machida}
}

\inst{
  Department of Physics, Okayama University, Okayama 700-8530
}

\recdate{}

\abst{
The stability of a  quantized vortex state in Bose-Einstein condensation
is examined within Bogoliubov theory 
for alkali atom gases confined in a harmonic 
potential under forced rotation.
By solving the non-linear Bogoliubov
equations coupled with the Gross-Pitavskii equation,
the elementary excitations and the total
energy of the systems are calculated as a function of rotation velocity.
There are two distinct criteria of vortex stability;
The position of the excitation energy levels 
relative to the condensate energy level
yields the local stability criterion, and
the total energy relative to that of
the non-vortex state
yields the global stability criterion.
The vortex stability phase diagram in the rotation velocity vs the
particle density of the system is obtained,
allowing one to locate the appropriate region
to observe the singly quantized vortex.
}

\kword{
 Bose-Einstein condensation, vortex, alkali atom gas,
 Gross-Pitaevskii equation, 
 Bogoliubov equation
}

\begin{document}

\sloppy
\maketitle


\section{Introduction}

There has been much attention focused on the Bose-Einstein 
condensation (BEC) experimentally and theoretically
since it was realized in dilute alkali atom gases
in 1995~\cite{firstRb,hulet,ketterle}. 
The condensates have atom number
typically $\sim O(10^6)$ for $^{23}$Na and $^{87}$Rb,
their BEC transition temperatures are in a range of $O(\mu {\rm K})$,
and they are usually confined magnetically
in a harmonic trap~\cite{myatt,otherexpreview,dalfovo}.

In these systems, it is natural to expect a quantized vortex 
which has been already much studied
 in a superconductor (Fermion) and $^4$He (Boson)
for a long time.~\cite{vinen,donnelly}
%
While various microscopic theories for a vortex in Fermionic systems 
such as BCS-Gor'kov or Bogoliubov-de Gennes theories are 
developed,~\cite{vinen}
the corresponding mean-field framework 
for a vortex in Bosonic systems originally due to Bogoliubov
has not fully examined yet.~\cite{bogoliubov,pitaevskii,fetter}
Therefore, it is urgent to examine 
whether or not the widely used Bogoliubov theory,
which has been quite successful so far for describing present BEC systems
without a vortex,~\cite{dalfovo} can also yield a stable vortex.

In connection with the present dilute BEC systems of alkali atom 
gases,
theoretical~\cite{sinha,gpvortex,dal-stri,dodd,rokhsar,laserkarl,gp2dtime,svi} 
and experimental studies on a vortex have just started.
So far there is no report to observe a vortex experimentally.
One possible reason of the difficulty in producing a vortex
is that gases are confined magnetically in a harmonic trap and 
this object is not so easy to manipulate it
under rotation as compared with
superfluid $^4$He in a bucket.
There are some theoretical proposals~\cite{laserkarl,gp2dtime} to 
realize the vortex state in this situation.
The persistent rotation may be realized by shining moving laser
on a large BEC system,
which is examined theoretically by Marzlin {\it et 
al.}~\cite{laserkarl}.

Previously, we examined the stability problem of
a vortex in BEC systems confined
by a rigid wall~\cite{isoshima1} and 
by a harmonic potential~\cite{isoshima2}
within the framework of Hartree-Fock Bogoliubov theory
both at $T$=0 and $T \neq 0$.
The conclusions we have reached are summarized as follows:
(1) At $T$=0 the vortex is unstable
in the non-selfconsistent Bogoliubov theory 
and the self-consistent Popov theory.~\cite{isoshima1,isoshima2}
(2) Above a certain temperature the vortex becomes
stabilized by the presence of the increased non-condensate
fraction localized at the vortex center,
which effectively acts as a pinning potential, preventing it 
from spiraling out.~\cite{isoshima2}
(3) Even at $T=0$, the external laser-induced potential at the 
vortex center
has a similar effect to stabilize the vortex.~\cite{isoshima2}
In these discussions, the stability of the vortex state is examined
by the positive sign of the lowest eigenvalue~\cite{isoshima2}
while its negative sign means the eigenvalue becomes lower
than the ground state energy.
We call this stability ``local stability''.
The question of this type is still under lively
discussions.~\cite{gpvortex,dal-stri,dodd,rokhsar,svi,fettermail}

As in $^{4}$He system  under persistent rotation
around a symmetry axis,~\cite{donnelly}
another stability criterion of the vortex state is 
to compare the total energy of the vortex system with that of the 
non-vortex system.
We call this stability ``global stability'',
considering the energy landscape of the configurational space.
Here we study these two types of the vortex stability problem simultaneously,
when the BEC system is kept in rotation at $T=0$.

It is expected that above a certain critical angular
velocity $\Omega_{\rm global}$, a vortex nucleates 
as in superfluid $^{4}$He system or as in the lower critical field $H_{c1}$
in a type II superconductor.
As we will see in detail,
the applied rotation tends to stabilize a vortex in general,
therefore the above local stability criterion 
of the vortex formation is also satisfied even at $T=0$,
which is characterized by another angular velocity $\Omega_{\rm local}$.
The purposes of this paper are to determine these characteristic 
angular velocities;
$\Omega_{\rm local}$ and $\Omega_{\rm global}$ and to examine
the relationship between them.
Because the previous study~\cite{isoshima1,isoshima2}
corresponds to $\Omega=0$ here,
 we will gain further understanding
even for the local stability problem through this study
for various $\Omega$'s.

To attack this problem,
we solve the eigenvalue equation of the Bogoliubov theory,
which is non-selfconsistent.
If it gives a negative eigenvalue under a given 
condensate spatial profile 
determined by the associated no-linear Schr\"{o}dinger type equation
(Gross-Pitaevskii equation, or GP),
the quantized vortex cannot be stable even for a self-consistent calculation.

The arrangement of the paper is as follows:
After giving a brief description of the Bogoliubov theory in \S 2,
the vortex stability problem in a system trapped harmonically is 
discussed in  \S 3.
The numerical procedure for solving the GP equation and the
eigenvalue equation is the same as before,
described in refs.~\citen{isoshima1} and \citen{isoshima2}.
The final section is devoted to discussions and summary.

\section{Formulation of Bogoliubov Equation Under Rotation}

\subsection{General formulation}

We start with a system of interacting Bosons.
Two particles at ${\bf r}$ and ${\bf r}^\prime$ 
interact with the potential $g\delta({\bf r}-{\bf r}^\prime)$
where $g$ is a positive (repulsive) constant
proportional to the s-wave scattering length $a$, namely
$g=4\pi \hbar^2a/ m$ ($m$ the particle mass).
The system is kept under rotation with the angular velocity 
$\omega$ by an external force.
The relevant hamiltonian $\hat{\rm H}_{\rm rot}$
with the extra rotation term can be written as:
\begin{equation}
  \hat{\rm H}_{\rm rot}
  =
  \hat{\rm H} - {\bf \omega}\cdot\int{\bf r}\times{\bf p}\rv \D{\bf r}
 \label{eq:h0}
\end{equation}
where
\begin{eqnarray}
  \hat{\rm H}
  &=&
  \int \D{\bf r}
    \hat{\Psi}^{\dagger}\rv \left\{
      -\frac{\hbar^2 \nabla^2}{2m} + V\rv - \mu
    \right\} \hat{\Psi}\rv \nonumber
\\&&
  + \frac{g}{2} 
  \!\! \int \!\! \D{\bf r} 
  \hat{\Psi}^\dagger \rv \hat{\Psi}^\dagger \rv
  \hat{\Psi}\rv \hat{\Psi}\rv  ,                     
  \label{eq:h1}
\end{eqnarray}
${\bf p}(\bf r)$ is the momentum operator,
 $\mu$ is the chemical potential,
and $V\rv$ is the confining potential.

Following the standard method,
we decompose the field operator $\hat{\Psi}$ as
\begin{eqnarray}
  \hat{\Psi}\rv
  &=&
  \PSI\rv + \phi\rv      \label{eq:deco}
\\
  \phi\rv
  &\equiv&
  \langle \hat{\Psi}\rv \rangle.
\end{eqnarray}
We substitute the above decomposition eq. (\ref{eq:deco}) into eq. (\ref{eq:h0})
and ignore the higher order terms such as 
$\PSD \PSI \PSI$, $\PSD \PSD \PSI$, and $\PSD \PSD \PSI \PSI$ terms.
Then we rewrite the operator $\PSI\rv$ as
\begin{equation}
  \PSI\rv = \sum_q
    \left[
       u_q\rv \eta _q - v_q^*\rv \eta _q^{\dagger}
    \right]   \label{eq:psieta}
\end{equation}
where $\eta_q$ ($\eta_q^{\dagger}$) is the annihilation (creation) 
operator 
and $u_q\rv$ and $v_q\rv$ are the wave functions,
and the subscript $q$ denotes the quantum number.

\subsection{Cylindrical system}

We consider a cylindrically symmetric system and
a vortex line, if exists,  passes through the center of a cylinder,
coinciding with the rotation axis.
We use the cylindrical coordinates:  ${\bf r} = (r,\theta ,z)$.
The system is trapped radially
by a harmonic potential
\begin{equation}
  V(r) = \frac{1}{2}m(2\pi \nu)^2 r^2 ,  \label{eq:potential}
\end{equation}

\noindent
and is periodic along the $z$-axis whose length is $L$.
We focus on the lowest eigenstates of the momentum along $z$-axis and
the quantum number ${\bf q}$ in eq.~(\ref{eq:psieta})
is written as $(q_r, q_{\theta})$, where
$q_r = 1, 2, 3, \cdots $ and  $q_\theta = 0, \pm 1, \pm 2, \cdots$.

The condensate wave function $\phi\rv $ is expressed as
\begin{equation}
  \phi(r,\theta ,z) = \phi(r) {\rm e}^{{\rm i}w\theta } 
\end{equation}
where $\phi(r) $ is a real function and 
$w$ is the winding number.
The case $w=0$ corresponds to a system without a vortex
and $w=1(w=2)$ corresponds to a system with a singly 
(doubly) quantized vortex.
The $w> 2$ case is not considered in this paper.
The phases of $u_q\rv$ and $v_q\rv$ can be written as 
\begin{eqnarray}
u_{q}\rv &=& u_{q}(r){\rm e}^{{\rm i}(q_{\theta } + w)\theta}
\label{eq:u1}
\\
v_{q}\rv &=& v_{q}(r){\rm e}^{{\rm i}(q_{\theta } - w)\theta }.
\label{eq:v1}
\end{eqnarray}
The condition that the first order term in $\PSI\rv$ 
of our hamiltonian vanish is 
\begin{eqnarray}
  \Bigl[
    -\frac{\hbar^2}{2m}\left\{
      \frac{{\D}^2}{{\D}r^2} +\frac{1}{r} \frac{\D}{{\D}r}
      - \frac{w^2}{r^2}
    \right\} - \mu   \nonumber
&&\\
    +V(r) + g\phi^2(r)
    + \hbar \omega w
  \Bigr]\phi(r)
  &=& 0.             \label{eq:gp}
\end{eqnarray}
This non-linear Schr\"{o}dinger type equation is called the 
Gross-Pitaevskii equation.
The condition that the hamiltonian be diagonalized gives 
the coupled eigenvalue equations for $u_q\rv$, and $v_q\rv$ whose
eigenvalue is $\varepsilon_q$:
\begin{eqnarray}
  \Bigl[
    -\frac{\hbar^2}{2m}\left\{
      \frac{\D^2}{\D r^2} + \frac{1}{r} \frac{\D}{\D r}
      -\frac{(w + q_\theta)^2}{r^2}
    \right\} - \mu          \nonumber
&&\\
    + V(r) + 2g\phi^2(r)
    + \hbar \omega (w + q_\theta )
  \Bigr]u_{q}(r)            \nonumber
&&\\
  -g \phi^2(r) v_{q}(r) 
  &=&
  \varepsilon _{q} u _{q} (r)    \label{eq:bg1}
\\
  \Bigl[
    -\frac{\hbar^2}{2m}\left\{
      \frac{\D^2}{\D r^2} + \frac{1}{r} \frac{\D}{\D r}
      - \frac{(w - q_\theta)^2}{r^2}
    \right\}  - \mu          \nonumber
&&\\
    + V(r) + 2g\phi^2(r)
    + \hbar \omega (w - q_\theta)
  \Bigr] v_{q}(r)          \nonumber
&&\\
  -g \phi^2(r) u_{q}(r) 
  &=&
  -\varepsilon _{q} v_{q}(r) .   \label{eq:bg2}
\end{eqnarray}
The normalization condition for $\phi\rv$, $u_q\rv$, and $v_q\rv$ are
\begin{equation}
  2\pi \int |\phi(r)|^{2} r \D r = \frac{N}{L}, \label{eq:gpnor}
\end{equation}
\begin{equation}
  2\pi \int \left\{
    u_p^*(r) u_q(r) - v_p^*(r) v_q(r) 
  \right\} r \D r
  = \frac{1}{L} \delta _{p,q}, \label{eq:bgnor}
\end{equation}
where $N$  is the total particle number.

The following symmetry relation should be noted:
\begin{equation}
  (u, v, \varepsilon) \mbox{ at } q_\theta = n
\Leftrightarrow
  (v, u, -\varepsilon) \mbox{ at } q_\theta = -n
\end{equation}
We determine the signs of $q_{\theta}$ and $\varepsilon$ by the
normalization condition eq.~(\ref{eq:bgnor}).
The set of the eqs.~(\ref{eq:gp}), (\ref{eq:bg1}),
 (\ref{eq:bg2}), (\ref{eq:gpnor}), and
(\ref{eq:bgnor}) constitute Bogoliubov theory of our system.

When this Bogoliubov theory is extended to a finite temperature
such as Popov approximation (see for example ref.~\citen{isoshima1}),
the eigenstates characterized by the eigenfunctions $u_q$
and $v_q$, and the eigenvalue $\varepsilon_q$ are
interpreted as the quasiparticles of the system 
which contribute to the total particle density
as $f(\varepsilon)(|u_q|^2 + |v_q|^2)$ with 
$f(\varepsilon)$ being the Bose distribution function.

The energy of the system is given by 
\begin{eqnarray}
  E
  &=&
  \langle \hat{H}_{\rm rot} \rangle + \mu N        \nonumber
\\
  &=&
  E_{\phi}
  + \sum_q ( \mu - \varepsilon_q )
    2\pi L \! \int |v_{q}(r)|^2 r\, \D r .    \label{eq:E}
\\
  E_{\phi}
  &=&
  2\pi L \! \int \! \biggl[
    -\frac{\hbar^2}{2m}
    \phi^{*}(r) \! \biggl(
      \frac{\D^2}{\D r^2} +\frac{1}{r} \frac{\D}{\D r} - \frac{w^2}{r^2}
    \biggr) \! \phi(r)           \nonumber
\\&&
    + V(r)|\phi(r)|^2
    + \frac{g}{2}|\phi(r)|^4
  \biggr] r\, \D r       
  - w \hbar \omega N.     \label{eq:E0}
\end{eqnarray}
Note that the coefficient of $\omega$ in eq.~(\ref{eq:E0}) is a constant.
Although $\varepsilon$'s and $v$'s are affected by the angular 
velocity  $\omega$,
these contributions to $E$ are negligibly small
and we will use $E_{\phi}$ as the total energy from now on.

\subsection{Normalization}

%

The condensate has $N$ particles in the length $L$
and the area density is $n_{z} = N/L$.
The mass of a particle is $m$ and the s-wave scattering length is $a$.
The length $r$ is measured as $r^{\prime} = (\sqrt{2mh\nu}/\hbar)r$.
We introduce the density unit
$n_{0} \equiv \sqrt{h\nu /g}$ where $g = 4\pi \hbar^{2}a/m$.
The density and energy are scaled by $n_{0}$ and $h\nu$
respectively.
Therefore, the normalized quantities are:
$\phi^\prime(r^\prime)  \equiv \frac{1}{\sqrt{n_{0}}}\phi(r)$, 
$u^\prime(r^\prime) \equiv  \frac{1}{\sqrt{n_{0}}}u(r)$,
$v^\prime(r^\prime) \equiv  \frac{1}{\sqrt{n_{0}}}v(r)$,
$\varepsilon^\prime_q\equiv \frac{1}{h\nu}\varepsilon_q$, and
$\mu^\prime \equiv \frac{1}{h\nu}\mu$.
The  angular velocity $\omega$ is also scaled by $2\pi\nu$ and the normalized
rotation is $\Omega \equiv \frac{\omega}{2\pi\nu}$.

Equation (\ref{eq:gp}) with eq. (\ref{eq:potential})
becomes in a dimensionless form
\begin{eqnarray}
  \Bigl[
    - \left\{
      \frac{\D^2}{\D r^{\prime 2}} +\frac{1}{r^{\prime}} \frac{\D}{\D r^{\prime}}
       - \frac{w^2}{r^{\prime 2}}
    \right\} - \mu ^{\prime}   \nonumber
&&\\
    + \frac{1}{4}r^{\prime 2} + \phi^{\prime 2}(r^{\prime})
    + w\Omega 
  \Bigr]\phi^{\prime}(r^{\prime})
  &=& 0.             \label{eq:gpN}
\end{eqnarray}
Equations (\ref{eq:bg1}) and (\ref{eq:bg2}) is transformed in a similar 
way.
Equations (\ref{eq:gpnor}) and (\ref{eq:bgnor}) are rewritten as
\begin{eqnarray}
  \int |\phi^{\prime}(r)|^{2} r^{\prime} \D r^{\prime} &=& 4a n_{z}.
\\
  \int \left\{
    u_p^{\prime *}(r^{\prime}) u_q^{\prime}(r^{\prime})
    - v_p^{\prime *}(r^{\prime}) v_q^{\prime}(r^{\prime}) 
  \right\} r^{\prime} \D r^{\prime}
  &=&
  \frac{4a}{L}\delta_{p,q}. \label{eq:bgnorN}
\end{eqnarray}
The energy $E_{\phi}$ is rewritten as
\begin{eqnarray}
  \frac{E_{\phi}}{h\nu N}
  &=&
  \frac{1}{4a n_{z}}
  \int  \biggl[ \! -\phi^{*\prime}(r^\prime) \!
    \biggl(
      \frac{\D^2}{\D r^{\prime 2}}
      +\frac{1}{r^{\prime}} \frac{\D}{\D r^{\prime}}
      -\frac{w^2}{r^{\prime 2}}
    \biggr) \! \phi^\prime(r^\prime)          \nonumber
\\&&
    + \frac{1}{4}r^{\prime 2}|\phi^\prime(r^\prime)|^2
    + |\phi^\prime(r^\prime)|^4
  \biggr] r^\prime \D r^\prime    
  - w \Omega.     \label{eq:E000}
\end{eqnarray}
Here $4a n_{z} $ and $\frac{4a}{L}$ are just dimensionless
numbers. Since the parameter $\frac{4a}{L}$ does not change the following 
results within the present Bogoliubov framework,
the system is characterized
by the number $a n_{z}$ and the normalized rotation
$\Omega = \frac{\omega}{2\pi \nu}$.

\section{Two Kinds of Stability}

We solve the coupled equations;
eqs. (\ref{eq:gp}), (\ref{eq:bg1}), (\ref{eq:bg2}), and (\ref{eq:bgnor})
for the gas of $^{23}$Na atoms trapped radially
by a harmonic potential eq. (\ref{eq:potential}).
The area density per unit length along the $z$-axis is chosen to be
$a n_z = 2.75 \sim 137.5$.
(When the scattering length is $a = 2.75 {\rm nm}$,
the area density $n_{z}$ varies from $0.1$ to $5$ ($10^4 /\mu {\rm m}$).)
We use the normalized number $\Omega$ to indicate
the angular velocity $\omega$.
$\Omega$ varies from $0$ to $1$.

\begin{figure}
  \begin{center}
    \leavevmode
    \epsfxsize=8cm
    \epsfbox{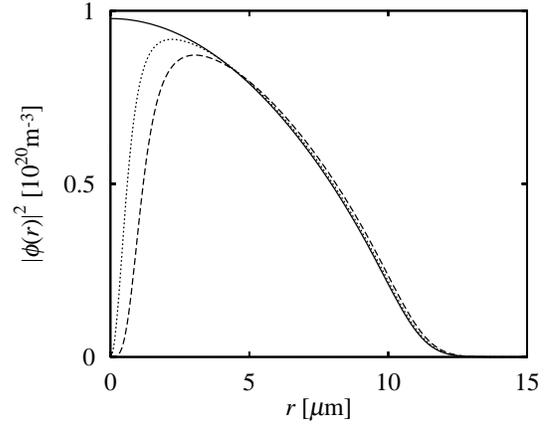}
  \end{center}
  \caption{
    The particle number density distribution $\phi^2(r)$
    along the radial direction. 
    Dashed (dotted) line is the singly (doubly) quantized 
    vortex case $w=1$ ($w=2$) while 
    solid line corresponds to the  non-vortex case.
    To draw this figure in actual units, we use the following parameters:
    the scattering length $a = 2.75 {\rm nm}$,
    $m = 3.81 \times 10^{-26} {\rm kg}$,
    the radial trapping frequency $\nu = 100 {\rm Hz}$,
    and $n_z = 2 \times 10^4 /\mu {\rm m}$.
  }
  \label{fig:dns}
\end{figure}

Since the GP equation eq.~(\ref{eq:gp}) is detached from the rest of
the equations,
it is easy to know the spatial profile of the condensate $\phi^2(r)$
by solving the GP equation.
Figure \ref{fig:dns} shows the typical density distribution of the 
system with and without the vortex along the radial direction.
The doubly quantized vortex $w=2$ case is also plotted
for comparison.

\subsection{Eigenvalues --- local stability}

When the lowest eigenvalue of the eigenvalue equations; 
eqs. (\ref{eq:bg1}) and (\ref{eq:bg2}) is negative in this 
formulation,
the fundamental assumption of BEC is broken down
as mentioned before.
In this framework, the eigenstate with negative eigenvalue means that
the system is unstable.
Suppose that
the negative eigenvalue appears in the vortex system
while it dose not appear in the corresponding non-vortex system.
Then the vortex state is unstable and prohibited
against the non-vortex system.
Dodd {\it et al.}~\cite{dodd} first consider the lowest eigenstate problem
in the vortex state under $\Omega=0$,
suggesting the presence of the negative 
eigenvalue for a harmonic potential.

\begin{figure}
  \begin{center}
    \leavevmode 
    \epsfxsize=8cm
    \epsfbox{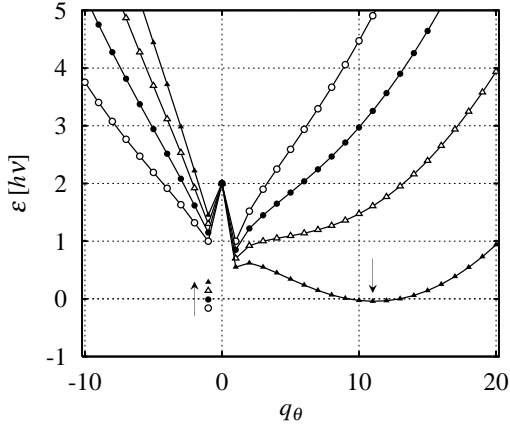}
  \end{center}
  \caption{
    The lowest edge of the eigenvalues along $q_\theta$ for selected
     angular velocities $\Omega$'s.
    Circles, filled circles, triangles, and filled triangles 
    correspond
    to $\Omega$ = 0, 0.15, 0.3, and 0.45 respectively.
    The eigenvalues at $q_\theta=-1$ move up while the eigenvalues
    in $q_\theta>0$ move down as $\Omega$ increases.
  }
  \label{fig:ene-wing}
\end{figure}

We perform here the extensive computations for solving the eigenvalue 
equations to know the lowest eigenvalue under various external 
angular velocities $\Omega$.
We show below the results of $a n_{z}=55$ as an example.
In Fig.~\ref{fig:ene-wing}, the lowest edge 
of the lowest eigenvalues,
above which other eigenvalues are densely distributed, is plotted as a function of the quantum 
number $q_{\theta}$ for several $\Omega$'s.
In the small $\Omega$ cases the lowest eigenvalue occurs
for the eigenstate characterized by the quantum number $q_\theta=-1$,
while in the larger $\Omega$ cases
some positive eigenvalue of $q_{\theta}$ around $q_{\theta}\sim 10$
becomes negative as $\Omega$ increases.
The eigenvalue at $q_\theta=0$ does not change with $\Omega$.

\begin{figure}
  \begin{center}
    \leavevmode 
    \epsfxsize=8cm
    \epsfbox{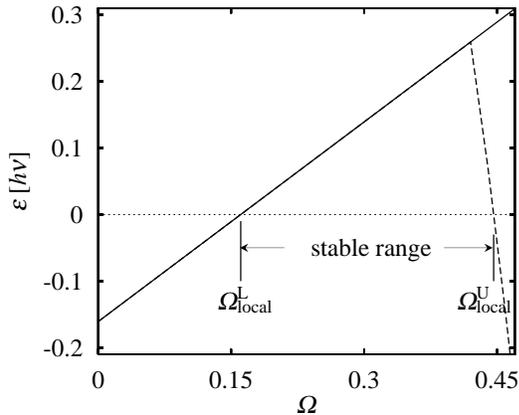}
  \end{center}
  \caption{
    The trace of the lowest eigenvalues as a function of $\Omega$. 
    Solid line corresponds to the eigenvalue of the 
    $(q_\theta, q_r) =(-1,1)$ state
    and dashed line is other states with the lowest 
    eigenvalue ($q_{r}=1$ and $q_{\theta}>0$). 
    For the stable region $0.16085 < \Omega < 0.44610$ all the
    eigenvalues in the system are positive.
  }
  \label{fig:ene-low}
\end{figure}

We determine the critical $\Omega$ where
the lowest $\varepsilon_{q}$ changes
its sign.
The trace of the 
lowest eigenvalues as a function of $\Omega$ is displayed
in Fig.~\ref{fig:ene-low}.
The solid straight line is the eigenvalues at $q_\theta=-1$
and the dashed line is ones in $q_\theta>0$.
The former becomes negative to positive at
$\Omega_{\rm local}^{\rm L} = 0.16085$
while the latter becomes negative at
$\Omega_{\rm local}^{\rm U}= 0.44610$. 
Therefore, the vortex state is locally stabilized in the region
$0.16085 < \Omega < 0.44610$ for the present example of $an_z=55$.
We call the stability defined by the sign of the lowest eigenvalue
 ``local stability''.
We define these critical $\Omega$'s as
$\Omega_{\rm local}^{\rm L} (=0.16085)$
and $\Omega_{\rm local}^{\rm U} (=0.44610)$.

\subsection{Interpretation of local stability}
 
The occurrence of this negative eigenvalue
at $q_\theta=-1$ may
correspond to the `spiraling out' of vortex predicted by 
Rokhsar~\cite{rokhsar}.
Strictly speaking,
the negative eigenvalues only means that
the system with a single vortex is unstable and prohibited.
But the spatial variation of the eigenfunctions $u_{q}$ and $v_{q}$ suggests
how the instability, which accompanies negative 
eigenvalue $\varepsilon_{q}$, actually occurs.
 
\begin{figure}
  \begin{center}
    \leavevmode 
    \epsfxsize=8cm
    \epsfbox{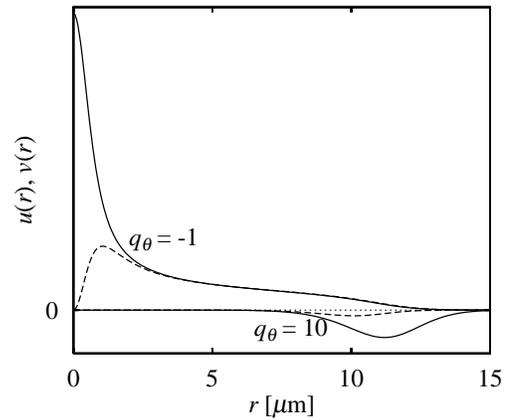}
  \end{center}
  \caption{Wavefunctions $u$ (solid line) and $v$ (dashed line)
  of the lowest modes at $q_{\theta}=-1$ (upper lines) and 
  at $q_{\theta}=10$ (lower lines).
  The system have $\Omega = 0$.
  These wavefunctions changes little
  even when $\Omega$ and the sign of $\varepsilon$ changes.
  The length unit is same as that of Fig.~\protect\ref{fig:dns}.
  }\label{fig:uv}
\end{figure}

\begin{figure}
  \begin{center}
    \leavevmode 
    \epsfxsize=8cm
    \epsfbox{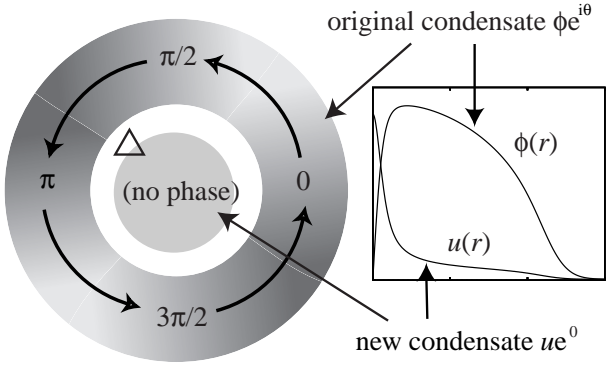}
  \end{center}
  \caption{
  The calculated wavefunctions $\phi(r)$ and $u(r)$ are showed in the
  right hand panel.
  They correspond to Figs.~\protect\ref{fig:dns} and \protect\ref{fig:uv}.
  The schematic figure of the phase change
  in the radial direction is drawn in the left hand panel.
  Because the new condensate without phase singularity
  appears at the center, the original vortex core (the triangle) must
  move outward.
  }
  \label{fig:localL}
\end{figure}

When the sign of the energy of
the lowest $q_{\theta}=-1$ state changes from positive to negative 
at $\Omega = \Omega_{\rm local}^{\rm L}$,
this (excited) state must become the condensate state.
Note that the original condensate was at $E=0$.
It is seen from Fig.~\ref{fig:uv} that
this $q_{\theta}=-1$ state is localized around $r=0$.
Let us consider the transformation process of the condensate.
We show the schematic diagram in Fig.~\ref{fig:localL}.
If the condensate is replaced by
$u(q_{\theta}=-1)$ around $r=0$
whose effective angular momentum is $w + q_{\theta}=0$
as seen from eq.~(\ref{eq:u1})
and still has the winding number $w=1$ at larger $r$ 
region (Fig.~\ref{fig:localL}), 
the vortex center as a phase singularity
($\triangle$ in Fig.~\ref{fig:localL}) 
must exist at the border
between the $w + q_{\theta}=0$ region which is localized around $r=0$
and the $w=1$ region which remains outside.
The axial symmetry is broken and
the vortex center slips out from $r=0$. 

\begin{figure}
  \begin{center}
    \leavevmode 
    \epsfxsize=8cm
    \epsfbox{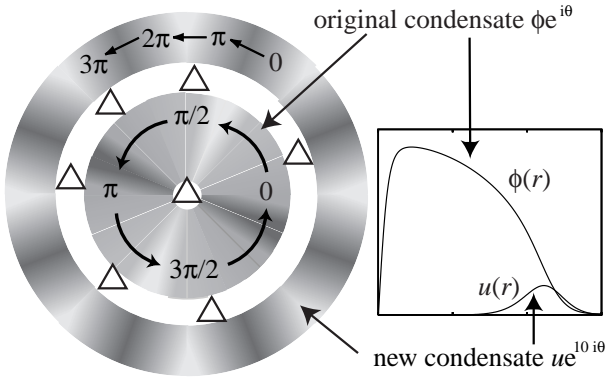}
  \end{center}
  \caption{
  The calculated wavefunctions $\phi(r)$ and $u(r)$ are shown in the 
  right hand panel.
  They correspond to Figs.~\protect\ref{fig:dns} and \protect\ref{fig:uv}.
  The schematic figure of the phase change
  in the radial direction is drawn in the left hand panel.
  As the new condensate at large $r$ has large ($\sim 10 \times 2\pi$ here)
  phase change, 
  many vortex cores (the triangle) must
  appear.
  }
  \label{fig:localU}
\end{figure}

On the other hand, 
at $\Omega = \Omega_{\rm local}^{\rm U}$,
$\varepsilon$ of one particular state with $q_{\theta}>0$,
say, $q_{\theta}=10$
becomes negative (see Fig.~\ref{fig:ene-wing}).
The lower curves of Fig.~\ref{fig:uv} are the wavefunctions for the
$q_{\theta}=10$ state.
They are localized at the edge region of the condensate $\phi$.
If the condensate has a large angular momentum
$w + q_{\theta} \simeq 10$ at large $r$
and has the angular momentum $w=1$ around $r=0$ (see Fig.~\ref{fig:localU}),
this means that many phase singularities (i.e. vortices)
must exist at the interface
between the $w + q_{\theta}\simeq 10$ region at large $r$
and the $w=1$ region at small $r$.
This transition stage will be followed by more complicated
transition process.


\subsection{Total energies --- global stability}

The comparison of the two energies 
in the systems with and without a vortex 
is another measure of the vortex stability.
The comparison indicates the global
stability of the vortex state relative to the non-vortex state.
We calculate the critical $\Omega$ in which 
the energy of the system with the quantized vortex is equal to 
the system without a vortex under the same condition.
We ignore the energy coming from the non-condensate part
 because the energy is 100 times smaller than 
the energy of the condensate in the present situation at $T = 0$.

In the non-vortex state, the energy $E_{w=0}$ is not 
affected by the rotation $\Omega$, as is seen from eq.~(\ref{eq:E0}).
The energy of the system without the vortex $E_{w=0}$, 
the energy of the singly quantized vortex system 
$E_{w=1}(\Omega)$, and
the energy of system with the doubly quantized vortex 
$E_{w=2}(\Omega)$ are evaluated from eq.~(\ref{eq:E000}) as
\begin{equation}
  E_{w =1}(\Omega) - E_{w =0}
  =
  h\nu N ( 0.22215 - \Omega)
\end{equation}
\begin{equation}
  E_{w =2}(\Omega) - E_{w =1}(\Omega)
  =
  h\nu N ( 0.42071 - \Omega).
\end{equation}
Therefore, the critical $\Omega_{\rm global}$ where the two energies
are equal; $E_{w=0}=E_{w=1}(\Omega)$ and 
$E_{w=1}(\Omega)=E_{w=2}(\Omega)$ are given by 
$\Omega_{\rm global}^{\rm L} = 0.22215$ and
$\Omega_{\rm global}^{\rm U}=0.42071$ for $a n_{z}=55$.
We call the  stability defined by the energy crossing
``global stability''.

\subsection{Two stabilities}

While the vortex stability in $^4$He systems is usually discussed in terms 
of the total energy (global stability in this paper),
the stability in the present dilute BEC systems is discussed~\cite{isoshima2}
 using the lowest eigenvalue (local stability in this paper).
These two stabilities are determined here and 
we can compare these quantitatively for a wide range of the
parameter values of $an_z$.

The local stability is satisfied for 
$0.16085 < \Omega < 0.44610$ and
the global stability is given by
$0.22215 < \Omega < 0.42071$ when $a n_{z}=55$.
In the substantially wide $\Omega$
region the vortex state of the rotating BEC systems is stable locally 
and globally.

\begin{figure}
  \begin{center}
    \leavevmode 
    \epsfxsize=8cm
    \epsfbox{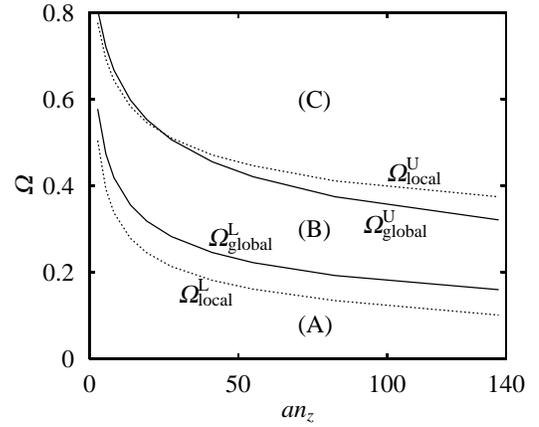}
  \end{center}
  \caption{
    Phase diagram in the $\Omega$ vs $an_z$ plane, showing
    the three characteristic regions: the vortex unstable region (A),
    the single vortex stable region (B) and
    the single vortex unstable region (C) where many vortices may appear.
    Solid lines show $\Omega_{\rm global}$ and dotted lines show 
    $\Omega_{\rm local}$.
    Upper two lines and lower two lines are $\Omega^{\rm U}$
     and $\Omega^{\rm L}$ respectively.
  }
  \label{fig:LandG}
\end{figure}

As mentioned previously, 
the system is characterized
by the number $a n_{z}$ and the rotation
$\Omega$.
In Fig.~\ref{fig:LandG} we show four critical $\Omega$'s
as a function of $a n_{z}$.
It is seen from Fig.~\ref{fig:LandG} that 
as $\Omega$ increases,
(1) the singly quantized 
vortex becomes stable locally first,
(2) the energy of this vortex state is lower than that in
the corresponding non-vortex state and the singly quantized vortex nucleates
in the system, corresponding to (B) in Fig.~\ref{fig:LandG}, and
(3) this single vortex state is kept stable up to either
$\Omega^{\rm U}_{\rm global}$
or $\Omega^{\rm U}_{\rm local}$.
These divide the parameter space into three regions:
the vortex unstable region (A),
the single vortex stable region (B) and
the single vortex unstable region (C) where many vortices may appear.

\section{Conclusion and Discussions}

In order to obtain more insights into the vortex stability problem
in the Bose-Einstein condensation of alkali atom gases
confined in a harmonic potential,
we have extended our previous work~\cite{isoshima1,isoshima2}
to the case where 
the cylindrical system is under forced rotation.

Our calculation is done within the framework of the non-selfconsistent
Bogoliubov theory and yields the vortex stability phase diagram shown in 
Fig.~\ref{fig:LandG}. The stability of the vortex state is examined by the
two different ways, that is,
the local stability and the global stability.
The previous cases~\cite{isoshima1,isoshima2},
which are consistent with the present results, correspond
to the $\Omega=0$ axis in this figure where the vortex is 
intrinsically unstable.
This instability is seen to occupy a finite region, rather than the
isolated line
confined at $\Omega =0$ for various densities $n_z$ and the scattering
lengths $a$.

We have examined not only the above intrinsic local stability, but also
the  global stability of the single vortex relative to the non-vortex state.
This allows us to draw the whole perspective for the vortex stability 
problem.
As is seen from Fig.~\ref{fig:LandG}, this global stability borderline
$\Omega_{\rm global}^{\rm L}$
corresponding to the vortex nucleation
is always situated above
the intrinsic stability borderline $\Omega_{\rm local}^{\rm L}$.
This means that for any $an_z$ values investigated
the singly quantized vortex, which is nucleated,
can exist as a (locally) stable object
under rotation at $T$=0.
It is to be noted that the present results in the non-selfconsistent
calculations
are applicable even for selfconsistent calculations such as the Popov
approximation.

We have performed a similar study for the rigid wall case instead of the 
present harmonic potential case. Our data show the same vortex 
stability phase diagram for the system sizes
up to 80$\times$(coherence length), confirming 
the previous conclusion~\cite{isoshima1} 
that the intrinsic vortex unstable region exists
at a lower $\Omega$ region.

Together with the previous finite temperature calculations~\cite{isoshima2},
the present calculation concludes that alkali atom Bose gases in BEC
can sustain and 
exhibit the stable vortex in appropriate temperatures and appropriate
rotations.

\section*{Acknowledgment}
The authors thank T. Ohmi for useful discussions.



\end{document}